\pdfoutput=1
\documentclass[11pt]{article}

\usepackage[T1]{fontenc}
\usepackage[utf8]{inputenc}

\usepackage[margin=1in]{geometry}
\usepackage{amsmath}
\usepackage{booktabs}
\usepackage{tabularx}
\usepackage[protrusion=true,expansion=false]{microtype}
\usepackage[hidelinks]{hyperref}
\hypersetup{
  pdftitle={Evidence Records for Public-Record Changes: Definitional Class Membership as the Certifiable Citation Unit},
  pdfauthor={Amadeus Brandes}
}

\title{Evidence Records for Public-Record Changes:\\[2pt]
\large Definitional Class Membership as the Certifiable Citation Unit}
\author{Amadeus Brandes\\[3pt]
\normalsize Independent Researcher, Germany\\
\normalsize \texttt{brandesamadeus@gmail.com}}
\date{September 2026}

\begin{document}
\maketitle

\begin{abstract}
Public records change in ways that matter for later interpretation, yet citing a change precisely is difficult: a reader must reconstruct prior versions, identify the changed fields, recover provenance, and---most consequentially---separate what the change demonstrably \emph{is} from what it is claimed to \emph{mean}. We argue these are two kinds of claim with two different epistemic statuses. Definitional class membership---whether a change matches an objectively checkable predicate, such as ``a primary endpoint field changed after the recorded primary-completion date''---is certifiable: membership can be independently verified by checking the predicate against the cited source. In our review design, importance or review-priority---whether a change is ``critical''---did not certify: competent reviewers did not converge on it. We introduce the \emph{Evidence Record}, a deterministic, hash-pinned, claim-bounded primitive that makes a public-record change citable through a record whose integrity any reader can verify, and that carries its definitional class memberships explicitly. Integrity is checkable from the published package alone; deciding most class predicates additionally requires registry state that records identify by hash but do not embed. We implement it in TrialDiff, a system for ClinicalTrials.gov protocol amendments, and report results in both directions of the boundary. On the certifiable side, deterministic event-classes over a fixed, non-random corpus of 100 breast-cancer-related ClinicalTrials.gov interventional trials yield citable records whose integrity is individually verifiable; our headline result shows that a naive ``primary endpoint changed after completion'' flag fires on 73 patches, but 63 co-occur with results posting and only 10 are clean---a separation the reconciliation-exclusion makes as a checkable definitional predicate, not as a validated priority judgment. Two defects found after publication, in opposite directions, were located by checking records against their own cited evidence and corrected by reissue rather than rewriting---the certifiable layer proving falsifiable in practice. On the uncertifiable side, a derived severity/priority label could not be certified under blinded adversarial review (reviewer confirmations of 4 to 17 of 30 against a pre-registered 24/30 gate), and we release it only as uncalibrated metadata. The certifiable citation unit for a public-record change is definitional class membership, not a priority ranking.
\end{abstract}

\section{Introduction}
\label{sec:intro}

Public records---clinical-trial registries, regulatory filings, public datasets---are revised over time, and those revisions frequently matter more than the records' initial state. A trial that changes its primary endpoint after data collection, a filing that removes a disclosure, a dataset that silently revises a figure: in each case the \emph{change} is the object of interest for later interpretation. Yet such changes are typically cited informally---reconstructed ad hoc from version histories, described in prose, and, critically, asserted to be important before any verifiable evidence object for the change exists.

The difficulty is not only mechanical. Recording a change reproducibly is, in principle, solved: content-addressed hashing~\cite{fips180}, canonical serialization~\cite{rfc8785}, provenance models~\cite{provdm}, and structured change formats~\cite{rfc6902} all exist. The harder problem is epistemic discipline---distinguishing what a change demonstrably \emph{is} from what it is claimed to \emph{mean}. These are two kinds of claim. The first is definitional: does the change match an objectively checkable predicate, such as the removal of a previously stated reason for stopping a terminated trial? The second is evaluative: is the change ``critical,'' ``high-priority,'' or indicative of a problem? We show, empirically and in both directions, that the first kind can be made certifiable here, while the second did not certify under our rubric-review design---and that conflating them is what makes change-citation untrustworthy.

We introduce the \textbf{Evidence Record}, a primitive that isolates the certifiable layer. An Evidence Record assigns a deterministic, generation-scoped identifier to a bounded public-record change event, carries the source's own change representation together with the set of changed paths, pins each cited source by content hash, records generator and ruleset versions, separates supported claims from explicit non-claims, and carries a deterministic set of definitional class memberships (\texttt{event\_classes}). The record is a pure function of its inputs: regenerating it from the same sources yields byte-identical canonical JSON and an identical record hash. It is claim-bounded: it asserts the structural fact of the change and nothing about intent, wrongdoing, compliance, or significance. Its verifiability is real but partial, and Section~\ref{sec:primitive} distinguishes the part a reader can check unaided from the part that still requires sources we do not package.

We instantiate the primitive in \textbf{TrialDiff}, a system that produces Evidence Records for ClinicalTrials.gov protocol amendments, and use it to test the boundary on a 100-trial corpus. Post-completion endpoint changes are a documented transparency concern. Florez et al.~\cite{florez}, in a cross-sectional study of 755 cancer phase~3 randomized trials, found that 145 trials (19.2\%) changed a primary endpoint after initiation, that 102 of those 145 (70.3\%) were not reported in the corresponding publication, and that among the 120 trials whose primary-endpoint changes were observable in registry version history, just over half (52.5\%) made the change after primary completion. The phenomenon has a long empirical record: selective reporting of outcomes relative to protocols~\cite{chan2004}, discrepancies between registered and published primary outcomes~\cite{mathieu2009}, and---closest in spirit to the present work---the COMPare study, which prospectively checked reported outcomes against pre-specified ones and published corrections in real time~\cite{compare2019}. What that literature establishes by expert reading, this paper asks how to make citable and mechanically checkable. The domain is therefore a fair test of whether such changes can be made \emph{citable}---verifiable and bounded---rather than merely flagged. Our contributions are: (1)~the Evidence Record primitive with its determinism and claim-boundary guarantees; (2)~a set of deterministic, individually verified event-classes over real registry amendments, including a reconciliation-exclusion result that separates 10 clean post-completion endpoint changes from 63 reconciliation-confounded ones; and (3)~a negative result---a derived priority label that could not be certified under blinded review---that empirically marks the limit of the certifiable layer.

\section{The Evidence Record Primitive}
\label{sec:primitive}

An Evidence Record is a deterministic, claim-bounded object that asserts a single bounded change in a public record, addressable by a generation-scoped identifier, integrity-verifiable by any reader, and checkable against its cited sources to the extent those sources are packaged or identified. The primitive is defined by its fields, three guarantees---determinism, provenance, and claim boundary---and one separation: definitional class membership versus uncalibrated triage.

\paragraph{Fields.} A record carries: \texttt{record\_id}, a deterministic identifier derived from the record's content, serving as the citation key and stable only within a generation; \texttt{source}, for each cited source a retrieval URL, a content hash at the observed version, and version reference(s) (in TrialDiff, the patch payload is carried directly; from/to version payloads are identified by stored snapshot hashes without being reproduced); \texttt{change}, a canonical representation of the change as the set of changed paths together with the source's own patch operations, each carrying an operation, a path, and the new value---prior values are not embedded, and live in the hash-identified from-version payload; \texttt{ruleset\_hash}, a hash of the logic version that produced any derived fields; \texttt{event\_classes}, a deterministic set of definitional class memberships (a record is one-per-source-change and may carry several classes; the primitive permits an empty set); \texttt{triage\_label}, a deterministic but uncalibrated classification, with \texttt{calibration\_status} recording the status of any external validation; \texttt{claims\_supported} and \texttt{claims\_not\_supported}, explicit lists of what the record does and does not assert; and generator/version metadata. The record's canonical JSON form is hashed to a canonical hash---the primitive's \texttt{record\_hash}---stored alongside the record, in the package manifest and the serving layer, rather than inside it: the hash of the canonical form cannot be a field of that form.

\paragraph{Determinism.} A record is a pure function of its cited source content, the source-selection rules, the canonicalization rules, the generator version, and the \texttt{ruleset\_hash}. The same inputs produce byte-identical canonical JSON and therefore an identical \texttt{record\_hash} and \texttt{record\_id}. Canonicalization---stable key ordering and fixed normalization, including a sorted \texttt{event\_classes} set---is what makes the hash reproducible. Regenerating a record from the same inputs is a verification, not a re-derivation.

\paragraph{Provenance and immutability.} Each source is pinned by content hash at the observed version, so the record is derivable from those sources alone; the record's own canonical form is hashed as \texttt{record\_hash}. Published records are immutable: a logic change produces new records under a new \texttt{ruleset\_hash} and never mutates the bytes of an existing record, which would break its hash and any citation to it. Verification is mechanical: retrieve each cited or archived source payload, confirm its content hash, recompute the canonical form, and confirm the \texttt{record\_hash}.

\paragraph{Integrity and decision fidelity.} Two distinct properties are often conflated under ``verifiable,'' and the primitive supports them unequally. \emph{Record integrity}---that a published record is the exact object it claims to be---is recomputable by any reader from the published package alone: each record file's SHA-256 equals its stored canonical hash, the manifest fixes every file in the package, and an identity digest over the sorted \texttt{event\_id}/canonical-hash listing summarises a whole generation in one value a reader can reproduce without access to the generating database. \emph{Predicate-decision fidelity}---that a class membership was decided correctly against the registry state---is not recomputable from packaged bytes for every membership. Four of the five predicates gate on from-version record state that the records do not embed: completion status, prior enrollment value, the content of a removed item, terminal status. Only the reconciliation class reads nothing but the patch. In the generation reported here, 80 of 109 memberships are decidable from the embedded patch alone; the remaining 29 require registry-sourced slices the package does not include.

The gap is bounded and named rather than silent. Each record carries \texttt{from\_snapshot\_hash} and, where stored, \texttt{to\_snapshot\_hash}, so the omitted source payloads are identified by content hash even where they are not reproduced: a reader knows exactly which payloads a decision depended on, and can confirm that a payload obtained elsewhere is the one used, without being able to reconstruct it from the package. Source closure in this sense is a property of predicate shape, not of packaging effort---a predicate that reads prior state will always require that state---and closing it requires shipping the registry-sourced input slices alongside the records, together with an independent checker that does not import the generating implementation. That work is not done here, and no claim of independent reconstruction is made. Replaying a generation and obtaining identical bytes establishes internal consistency of ingestion and derivation, not independent fidelity to ClinicalTrials.gov.

\paragraph{Claim boundary.} Every record states \texttt{claims\_supported} and \texttt{claims\_not\_supported}. A record asserts only the structural fact of the change as represented in its cited source; it does not assert intent, wrongdoing, regulatory compliance, disclosure status in any other venue, or substantive significance. The \texttt{triage\_label} is uncalibrated and is explicitly neither a validated priority nor a finding. This separation---definitional \texttt{event\_classes} as the certifiable layer, \texttt{triage\_label} as uncalibrated metadata---is the primitive's central discipline and the subject of the results that follow.

\section{TrialDiff Implementation}
\label{sec:impl}

TrialDiff instantiates the primitive for ClinicalTrials.gov protocol amendments~\cite{ctgov,zarin2011,zarin2016}. Each registered trial has a sequence of versioned records; TrialDiff retrieves the version history, pins each version's payload by content hash, and obtains the registry's own patch for each adjacent version pair.

\paragraph{Sources and change representation.} The registry itself serves the change between adjacent versions as a JSON~Patch document~\cite{rfc6902}; TrialDiff retrieves it from the version-history interface, pins it by content hash both as served and in canonical form, and derives from it the set of changed paths, expressed as JSON Pointers~\cite{rfc6901}. Patch operations carry an operation, a path, and the new value; prior values are not embedded in the record, and live in the from-version payload, which the record identifies by hash. For a trial \texttt{nct\_id}, TrialDiff stores version payloads locally, hashes each stored from- and to-version record, and carries those hashes in the Evidence Record as \texttt{from\_snapshot\_hash} and \texttt{to\_snapshot\_hash}. The patch payload itself is stored in the record, with its source URL and hashes. Verification therefore checks the patch and stored snapshot hashes against the retrieved or archived payloads, rather than relying on a live endpoint alone. Where a to-version snapshot was not stored, the record carries the patch and the from-version hash but no to-version hash; predicates that read to-version state must treat that absence as unknown rather than as an absent field, a distinction whose violation produced the first of the two corrections reported in Section~\ref{sec:errata}.

\paragraph{Records.} Generation is one record per patch. A record's identifier has the form \texttt{evt\_\{nct\_id\}\_\allowbreak v\{from\}\_\allowbreak v\{to\}\_\allowbreak\{hash\}}, where \texttt{\{hash\}} is a prefix of the SHA-256 over the record's canonical content; this binds the identifier to both the patch coordinates and its content. Field names follow the implementation: the primitive's \texttt{record\_id} is realised as \texttt{event\_id}; the canonical hash is stored outside the record, in the package manifest, the serving database, and the HTTP entity tag; and \texttt{ruleset\_hash} is realised as \texttt{event\_class\_rule\_set\_hash}, alongside a combined hash that also covers the triage logic. The record's \texttt{change}, snapshot and patch hashes, \texttt{event\_classes}, \texttt{triage\_label}, and claims are serialized to canonical JSON with stable key ordering and a sorted \texttt{event\_classes} set, and the canonical form is hashed to the record's canonical hash. Two independent regenerations over the working corpus produced byte-identical payloads and hashes, confirming the determinism guarantee holds with the \texttt{event\_classes} field present.

\paragraph{Event-classes.} A patch's \texttt{event\_classes} is the set of definitional predicates it satisfies (the five predicates are specified in Section~\ref{sec:results}). Generation is membership-driven: TrialDiff emits a record only for patches that satisfy at least one predicate, and a patch satisfying several predicates yields a single record carrying all of them---keeping \texttt{event\_id} and canonical hash in one-to-one correspondence with the underlying patch.

\paragraph{Triage label.} Each record also carries a deterministic \texttt{triage\_label}, retained from an earlier severity classifier, with \texttt{calibration\_status} set to \texttt{uncalibrated}. The label is reproducible but is not a validated review priority; the attempt to certify it, and the decision to retain it only as uncalibrated metadata, are reported in Section~\ref{sec:results}. A record's \texttt{claims\_supported} enumerate the satisfied event-class predicates, while \texttt{claims\_not\_supported} include the reconciliation non-claim---co-occurrence with results posting is recorded, harmlessness is not asserted.

\section{Results}
\label{sec:results}

\subsection{Corpus and generation}

The corpus is 100 ClinicalTrials.gov interventional trials (breast-cancer-related), comprising 4{,}485 adjacent-version patches. Evidence Record generation is membership-driven: a record is emitted for each patch that satisfies at least one event-class predicate, carrying all satisfied classes as a sorted set. All counts reported below are properties of one generation of the record set, identified by its event-class rule-set hash; identifiers and hashes rotate when that hash changes, and a citation resolves to the generation it names. In the generation reported here (rule-set hash \texttt{74a6f55a}\ldots, deposited as v0.1.3), generation yields 97 records across 54 trials---85 single-class and 12 two-class, with no record carrying three classes (109 class memberships in total). Generation is deterministic---the reproducible-builds discipline~\cite{lamb2022} applied to a data artifact: independent regenerations produce byte-identical canonical payloads and record hashes, with \texttt{event\_classes} sorted before serialization, and the published package carries an identity digest over the sorted record listing that a reader can recompute from the package alone. The corpus is not a random sample of the registry, so counts are illustrative of the method rather than population estimates.

\subsection{Certifiable event-classes}

We define five event-classes as deterministic predicates over the patch and the from-version state. Each is a membership fact---an objectively checkable condition---not a judgment, so verification reduces to confirming, against the cited source, that the predicate fired correctly. Counts over the corpus:

\begin{center}
\begin{tabularx}{\textwidth}{Xr}
\toprule
Event-class & Count \\
\midrule
Primary endpoint changed after primary completion, reconciliation excluded & 10 \\
Secondary outcome item removed after primary completion & 12 \\
Enrollment changed to zero & 3 \\
\texttt{whyStopped} explanation removed in terminal context & 4 \\
Outcome edit co-occurring with results posting (reconciliation) & 80 \\
\bottomrule
\end{tabularx}
\end{center}

Because membership is definitional rather than evaluative, the claim attached to a published generation is bounded exhaustiveness: within the named corpus and under the named rule-set hash, the published memberships are the complete set the stated predicates select. This is a stronger and more useful claim than precision alone, and it is falsifiable in both directions---a member that does not satisfy its predicate, and a non-member that does, are each errata against a fixed claim rather than matters of undocumented selection. Both directions have in fact occurred, and both are reported in Section~\ref{sec:errata}. The claim does not extend beyond the named corpus or to any population rate.

The predicates are correspondingly literal. The secondary-outcome-removal predicate, for example, requires that the removed item's normalised content be absent from the to-version list, which excludes list-reindex artifacts in which an item is renumbered rather than removed; an earlier formulation that omitted this requirement returned members of exactly that kind.

\subsection{Headline: the reconciliation boundary}

The class definitions matter because a naive predicate over-reports. An inclusive ``primary endpoint field changed after primary completion'' flag fires on 73 patches. Of these, 63 co-occur with results posting (a change to \texttt{hasResults} or the results section) and only 10 do not. The 63 are \emph{confounded} by results reconciliation: when results are first posted, outcome fields are frequently edited as part of that posting, so a post-completion endpoint edit co-occurring with results posting cannot be cleanly attributed as a substantive protocol change. A definitional reconciliation-exclusion separates the 10 clean cases from the 63 confounded ones deterministically and reproducibly. What certifies here is the predicate, not a priority judgment: a score could encode the same rule, but it would then be certifying definitional membership, not the importance ranking the score purports to supply---and the ranking is what does not certify (see Section~\ref{sec:negative}).

The chain is recomputed for each generation rather than carried forward, and the partition---73 inclusive flags resolving into 63 reconciliation-confounded and 10 clean---is recorded in the published package's determinism attestation for the generation reported here.

We state the boundary carefully. The reconciliation class asserts co-occurrence with a results signal, not harmlessness: the 63 are not proven benign, only structurally confounded, and the record's claims make this explicit. The exclusion removes 86\% of the naive flag's volume on this corpus; the generalizable claim is the \emph{need} for a reconciliation class, not the specific ratio.

\subsection{Multi-class records}

A single change can satisfy several definitional predicates, and the record carries them as a set rather than being duplicated. In the generation reported here, 12 records carry two classes and none carries three. The worked example is NCT05415215~v29$\to$v30, record \texttt{evt\_NCT05415215\_v29\_v30\_\allowbreak d1ea363dbff7} (canonical hash \texttt{34f78893}\ldots), which is simultaneously a secondary-outcome removal after primary completion and an outcome edit co-occurring with results posting. Its \texttt{claims\_supported} lists each class's predicate; its \texttt{claims\_not\_supported} includes the reconciliation non-claim. The record is one-per-source-change, keeping the identifier and hash in one-to-one correspondence with the underlying patch. Which records carry multiple classes is a property of the generation, not of the registry: the corrections in Section~\ref{sec:errata} changed both the membership set and the overlap distribution.

\subsection{The uncertifiable side: priority calibration}
\label{sec:negative}

The same system originally derived a severity label intended as a review-priority signal. We attempted to certify it as a validated priority standard under blinded adversarial review: reviewers assigned priority from raw change material with the classifier's output hidden, against a pre-registered gate requiring at least 24 of 30 critical-tier records to be confirmed (critical false positives under 20\%). Across two rule-tightening cycles and four fresh reviewers, critical confirmations were 4, 5, 12, and 17 of 30 on identical records under the same rubric. No reviewer cleared the gate, and the fourfold spread across reviewers indicates the deeper problem: the target is not stable. A classifier cannot be calibrated to a reference that competent independent judges do not converge on. We report confirmation counts against the pre-registered gate rather than a chance-corrected agreement coefficient~\cite{cohen1960}, because the question was whether any reviewer confirmed the classifier's critical tier at the required rate, not how far reviewers agreed with one another; the spread across reviewers is reported as the observation it is. We therefore retain severity only as uncalibrated triage metadata, flagged as such in every record, and treat definitional class membership---not priority scoring---as the certifiable layer. The negative is bounded: it concerns this construct under this review design, not a general claim that priority classification is impossible.

\subsection{Corrections under the claim boundary}
\label{sec:errata}

Two defects have been found in published generations of the record set and corrected. Both were found by checking records against their cited sources---the operation the claim boundary is designed to make possible---and both were corrected by generating a new record set under a new rule-set hash rather than by editing published bytes.

\paragraph{E1: a storage gap reported as a registry change.} In the first published generations, the \texttt{why\_stopped\_removed\_in\_terminal\_context} predicate asserted a change that the records' own patches contradicted. The predicate reads the to-version state to establish that a previously stated \texttt{whyStopped} value is now absent; where the to-version snapshot had not been stored, it read the missing snapshot as an absent field. Nine of the thirteen records carrying the class contained no patch operation touching \texttt{whyStopped} at all: the field did not change between those versions. The defect was detectable precisely because each record embeds the registry's own patch alongside its \texttt{claims\_supported}, so the assertion and its evidence sit in the same object and can be compared without reference to the generating system. These were false positives, and the corrected generation retracts them.

\paragraph{E4: sequential index resolution and a path-shape gap.} In a later generation, the \texttt{secondary\_outcome\_removed\_after\_primary\_completion} predicate resolved JSON~Patch removal targets against the original from-version array. JSON~Patch array indices are evaluated sequentially~\cite{rfc6902}---after an item at index~$i$ is removed, the following item occupies index~$i$---so a patch containing repeated removals at one path was examined at its first target repeatedly and its later targets not at all. A related gap: operations on the whole \texttt{secondaryOutcomes} array fell outside the predicate's path-shape filter, so the wholesale deletion of every secondary outcome was invisible to the class. The defect was found by re-scanning all 4{,}485 adjacent-version patches with a candidate filter implemented independently of the production predicate, together with a coverage invariant---any post-completion patch that shrinks the secondary-outcome list must be recognised as a structural candidate---to catch shapes the filter itself might miss. Three memberships disagreed with the published set: \texttt{NCT01224678}~v109$\to$v110 and \texttt{NCT03094169}~v11$\to$v12 (whole-array removal), and \texttt{NCT03734029}~v29$\to$v30 (sequential removals at one index). All three were false negatives; every published positive remained valid. E4 adds memberships and retracts none.

\paragraph{Direction of error, and what it means for a citation.} The two corrections differ in the way that matters to anyone who has already cited a generation. E1 retracted nine memberships: a citation to a \texttt{whyStopped} membership in an affected generation may no longer stand, and the reader must check it against the erratum. E4 retracted none and added three: a citation to a generation affected only by E4 still stands, and the correction means the published set was under-inclusive rather than wrong. Recording which direction a correction runs is therefore not bookkeeping but the decision-relevant fact, and each affected deposit carries a notice pointing both to its immediate successor---for correction provenance---and to the concept identifier, for current discovery.

\paragraph{What the corrections demonstrate.} Three properties held under real defect conditions. First, the claim boundary made both defects findable: a record that states what it asserts, beside the source evidence for that assertion, can be checked and can therefore be shown to be wrong. An importance ranking admits no equivalent check---there is no source against which ``critical'' is falsified---which is the practical form of the boundary this paper reports. Second, hashing normalised implementation source bytes rather than the class-definition text alone proved load-bearing. The defect corrected in E4 lay in the implementation, not in the stated rule: the published definition already required that a removed outcome not reappear elsewhere in the to-version list, and the code failed to implement it. A correction of that shape need not alter the definition text at all, and under an identity derived from definitions alone the corrected records could have shipped indistinguishable from their predecessors. Pinning implementation bytes makes rotation unconditional. Third, immutability held. No published package was rewritten and no deposit withdrawn; each correction shipped as a new generation under a new rule-set hash, with the erratum attached to the generations it affects.

\section{Discussion}
\label{sec:discussion}

The contribution is a verifiable, claim-bounded citation layer for public-record changes, not a universal classifier of their importance. What an Evidence Record certifies is definitional: that a change matches a stated, mechanically checkable predicate, pinned to hashed sources and reproducible from them. The derived \texttt{triage\_label} is retained only as uncalibrated metadata, and the failure to certify it under blinded rubric review is reported as the boundary of the certifiable layer, not as a defect to be tuned away. The constructive consequence is the paper's organizing claim: cite definitional class membership, not a priority score.

The results carry specific limits. The published generation claims bounded exhaustiveness only within the named corpus and under the named rule-set hash: it is a complete census of what the five stated predicates select in that fixed corpus, not registry-wide recall. The counts come from a single, non-random, breast-cancer-related corpus, so they illustrate the method rather than estimate population rates; in particular, the reconciliation result generalizes as the \emph{need} for a reconciliation-exclusion class, not as the 86\% ratio observed here. The five event-classes are a starter set chosen to exercise the mechanism, not an exhaustive taxonomy of registry changes; the primitive permits an empty class set, and the instance is extensible to further predicates. The negative result concerns one priority construct under one rubric-review design.

The mechanism reuses established components---content-addressed hashing, provenance pinning, structural diffing---and claims no novelty in them. The contribution is the discipline imposed on top: explicit claim boundaries on every record, and the separation of definitional class membership (certifiable, reproducible, citable) from importance judgments (which did not certify here). The closest prior framing is data citation for evolving data. The FORCE11 principles establish that data be citable through persistent identifiers with provenance~\cite{force11dcp}; the RDA Working Group on Data Citation recommends citing subsets of dynamic data by versioned query, timestamp, and result-set hash~\cite{rda2016}; both sit within the FAIR programme~\cite{fair2016}. Those recommendations cite a \emph{state} of a dataset. An Evidence Record cites a \emph{change} as a first-class object with content-addressed identity and an explicit claim boundary---a different citation unit, not a competing scheme. Software citation principles~\cite{swcite2016} face the analogous versioned-identity problem for code, and the generation-scoped identifiers here follow the same logic. Two further contrasts sharpen what this adds. Record-integrity regimes---tamper-evident logs, and regulated audit-trail requirements such as 21~CFR Part~11---establish that a stored record was not altered; they do not address the distinction drawn here, between a change's certifiable definitional membership and an un-certifiable importance label attached to it. And validating a generation pipeline once does not substitute for per-record verifiability: the guarantee here is that each record is independently re-derivable from its cited sources, so verification rests on the record itself rather than on continued trust in the generator. Beyond clinical-trial registries, the same primitive applies wherever a change in a public record must be cited and independently verified---regulatory filings, revisions to public datasets---though that generality is asserted by construction, not demonstrated here.

\section{Availability}
\label{sec:availability}

The record set is deposited as a citable archive with a concept identifier that always resolves to the current generation, and a version identifier for each generation. Cite the concept identifier for the record set as an ongoing object and the version identifier when exact counts matter: the counts and identifiers reported in this paper belong to the generation deposited as v0.1.3 under event-class rule-set hash \texttt{74a6f55a}\ldots (Brandes~\cite{zenodo}; concept identifier \href{https://doi.org/10.5281/zenodo.20801956}{10.5281/zenodo.20801956}). Superseded generations remain published and immutable, each carrying a notice naming its successor, so a citation to an earlier generation continues to resolve to the bytes it cited. Source code, the event-class predicates, the deterministic generation pipeline, the Evidence Record primitive specification, the errata, and the determinism attestation are in the tagged release corresponding to that generation.\footnote{\url{https://github.com/AMBRA7592/trialdiff/releases/tag/event-class-v0.1.3}, frozen at commit \texttt{bad85f87}. The deposited archive and the release asset are byte-identical.} The manifest fixes every file in the package, and \texttt{shasum -a 256 -c MANIFEST.sha256} verifies the package offline.

A live demonstration serves the record set at \url{https://trialdiff.vercel.app}. Two properties of that deployment are relevant to citation and were observed in production rather than merely intended. First, an identifier published in a superseded generation continues to resolve: it returns the exact immutable bytes it always returned, with no redirect and no substitution, while the response advertises its successor out of band through a link relation and a supersession index. Generations coexist and resolve independently. Second, each canonical record response carries an entity tag equal to the SHA-256 of the served bytes, which is also the record's canonical hash and its entry in the package manifest, so a reader can verify a cited record with a single conditional request rather than by downloading and re-hashing it. One caveat follows from serving canonical records as immutable: a cached response may carry status metadata captured before a generation changed, while its bytes and entity tag remain correct. The supersession index, which is not served as immutable, is the authoritative surface for current status.

\end{document}